\documentclass[12pt]{article}
\usepackage{amsfonts}
\usepackage{amssymb}
\usepackage{graphics,psboxit,amsmath}
\usepackage{subfigure}
\usepackage{graphicx}
\usepackage{verbatim}
\usepackage{epic}

\def\hybrid{\topmargin 0pt \oddsidemargin 0pt 
        \headheight 0pt \headsep 0pt
        \textwidth 16,5cm 
        \textheight 23,5cm 
        \marginparwidth .875in
        \parskip 5pt plus 1pt \jot = 1.5ex}

\hybrid

\def\baselinestretch{1.2}

\catcode`\@=11

\def\marginnote#1{}
\newcount\hour
\newcount\minute
\newtoks\amorpm
\hour=\time\divide\hour by60
\minute=\time{\multiply\hour by60 \global\advance\minute by-\hour}
\edef\standardtime{{\ifnum\hour<12 \global\amorpm={am}%
        \else\global\amorpm={pm}\advance\hour by-12 \fi
        \ifnum\hour=0 \hour=12 \fi
        \number\hour:\ifnum\minute<10 0\fi\number\minute\the\amorpm}}
\edef\militarytime{\number\hour:\ifnum\minute<10 0\fi\number\minute}

\def\draftlabel#1{{\@bsphack\if@filesw {\let\thepage\relax
   \xdef\@gtempa{\write\@auxout{\string
      \newlabel{#1}{{\@currentlabel}{\thepage}}}}}\@gtempa
   \if@nobreak \ifvmode\nobreak\fi\fi\fi\@esphack}
        \gdef\@eqnlabel{#1}}
\def\@eqnlabel{}
\def\@vacuum{}
\def\draftmarginnote#1{\marginpar{\raggedright\scriptsize\tt#1}}

\def\draft{\oddsidemargin -.5truein
        \def\@oddfoot{\sl preliminary draft \hfil
        \rm\thepage\hfil\sl\today\quad\militarytime}
        \let\@evenfoot\@oddfoot \overfullrule 3pt
        \let\label=\draftlabel
        \let\marginnote=\draftmarginnote
   \def\@eqnnum{(\theequation)\rlap{\kern\marginparsep\tt\@eqnlabel}%
\global\let\@eqnlabel\@vacuum} }

\def\draft2{
        \def\@oddfoot{\sl preliminary draft \hfil
        \rm\thepage\hfil\sl\today\quad\militarytime}
        \let\@evenfoot\@oddfoot \overfullrule 3pt
        \let\label=\draftlabel
        \let\marginnote=\draftmarginnote
   \def\@eqnnum{(\theequation)\rlap{\kern\marginparsep\tt\@eqnlabel}%
\global\let\@eqnlabel\@vacuum} }

\def\preprint{\twocolumn\sloppy\flushbottom\parindent 2em
        \leftmargini 2em\leftmarginv .5em\leftmarginvi .5em
        \oddsidemargin -.5in \evensidemargin -.5in
        \columnsep .4in \footheight 0pt
        \textwidth 10.in \topmargin -.4in
        \headheight 12pt \topskip .4in
        \textheight 6.9in \footskip 0pt
        \def\@oddhead{\thepage\hfil\addtocounter{page}{1}\thepage}
        \let\@evenhead\@oddhead \def\@oddfoot{} \def\@evenfoot{} }

\def\numberbysection{\@addtoreset{equation}{section}
        \def\theequation{\thesection.\arabic{equation}}}

\def\underline#1{\relax\ifmmode\@@underline#1\else
        $\@@underline{\hbox{#1}}$\relax\fi}

\def\titlepage{\@restonecolfalse\if@twocolumn\@restonecoltrue\onecolumn
     \else \newpage \fi \thispagestyle{empty}\c@page\z@
        \def\thefootnote{\fnsymbol{footnote}} }

\def\endtitlepage{\if@restonecol\twocolumn \else \newpage \fi
        \def\thefootnote{\arabic{footnote}}
        \setcounter{footnote}{0}} 

\catcode`@=12
\relax

\def\figcap{\section*{Figure Captions\markboth
        {FIGURECAPTIONS}{FIGURECAPTIONS}}\list
        {Figure \arabic{enumi}:\hfill}{\settowidth\labelwidth{Figure
999:}
        \leftmargin\labelwidth
        \advance\leftmargin\labelsep\usecounter{enumi}}}
 \relax
\def\tablecap{\section*{Table Captions\markboth
        {TABLECAPTIONS}{TABLECAPTIONS}}\list
        {Table \arabic{enumi}:\hfill}{\settowidth\labelwidth{Table
999:}
        \leftmargin\labelwidth
        \advance\leftmargin\labelsep\usecounter{enumi}}}
 \relax
\def\reflist{\section*{References\markboth
        {REFLIST}{REFLIST}}\list
        {[\arabic{enumi}]\hfill}{\settowidth\labelwidth{[999]}
        \leftmargin\labelwidth
        \advance\leftmargin\labelsep\usecounter{enumi}}}
 \relax
\makeatletter
\newcounter{pubctr}
\def\publist{\@ifnextchar[{\@publist}{\@@publist}}
\def\@publist[#1]{\list
        {[\arabic{pubctr}]\hfill}{\settowidth\labelwidth{[999]}
        \leftmargin\labelwidth
        \advance\leftmargin\labelsep
        \@nmbrlisttrue\def\@listctr{pubctr}
        \setcounter{pubctr}{#1}\addtocounter{pubctr}{-1}}}
\def\@@publist{\list
        {[\arabic{pubctr}]\hfill}{\settowidth\labelwidth{[999]}
        \leftmargin\labelwidth
        \advance\leftmargin\labelsep
        \@nmbrlisttrue\def\@listctr{pubctr}}}
 \relax
\makeatother

\def\ba{\begin{equation}}
\def\ea{\end{equation}}

\def\d{\delta}

\def\th{\theta}

\def\l{\lambda}

\def\s{\sigma}

\def\no{\noindent}

\def\qq{\qquad}

\def\IR{\relax{\rm I\kern-.18em R}}

\def \ha {{1\over 2}}

\begin{document}


\renewcommand{\theequation}{\thesection.\arabic{equation}}
\csname @addtoreset\endcsname{equation}{section}

\newcommand{\eqn}[1]{(\ref{#1})}
\newcommand{\be}{\begin{eqnarray}}
\newcommand{\ee}{\end{eqnarray}}
\newcommand{\non}{\nonumber}
\begin{titlepage}
\strut\hfill
\vskip 1.3cm
\begin{center}


{\large \bf Systematic classical continuum limits of integrable spin chains\\ and
 emerging novel dualities }

\vskip 0.5in

{\bf Jean Avan$^{a}$, Anastasia Doikou$^{b}$
 and Konstadinos Sfetsos$^{b}$}\\
\vskip 0.3in

{\footnotesize $^a$ LPTM, Universite de Cergy-Pontoise (CNRS UMR 8089),\\
F-95302 Cergy-Pontoise, France}\\

\vskip .1in

{\footnotesize
$^{b}$ Department of Engineering Sciences, University of Patras,\\
GR-26500 Patras, Greece}\\

\vskip .2in


{\footnotesize {\tt avan@u-cergy.fr, adoikou$@$upatras.gr, sfetsos@upatras.gr}}\\

\end{center}

\vskip .6in

\centerline{\bf Abstract}

\no
We examine certain classical continuum long wave-length limits
of prototype integrable quantum spin chains. We define the
corresponding construction of classical continuum Lax operators. Our discussion
starts with the XXX chain, the
anisotropic Heisenberg model and their generalizations and extends to the generic
isotropic and anisotropic $gl_n$ magnets.
Certain classical and quantum integrable models emerging
from special ``dualities'' of quantum spin chains, parametrized
by $c$-number matrices, are also presented.

\no

\vfill
\no


\end{titlepage}
\vfill
\eject


\tableofcontents

\def\baselinestretch{1.2}
\baselineskip 20 pt
\no

\section{Introduction}

Locally interacting discrete integrable spin chains have been
the subject of much interest since they cropped
up in string theory in the study of the AdS/CFT correspondence \cite{miha}.
Their classical, long wavelength limit provides a connection
to continuous $\s$-models describing
particular dynamics of the string (references on this subject can be found in e.g. \cite{fradkin, string}).

\no
This has lead us to tackle here the problem of formulating the classical continuum long wavelength limit of the
(simpler) quantum integrable closed spin chains in a way that directly preserves integrability. Accordingly
we will describe the classical Lax-matrix formulation, including the associated classical $r$-matrix
structure, which consistently yields the classical, long-wavelength limit,
derived for integrable closed quantum spin chain models (see e.g. \cite{fradkin, string}).

\no
We shall first describe and implement in detail the general Hamiltonian procedure.
Then we will tackle a number of specific examples, and explicitly compare
with already known results from alternative derivations. These identifications
will establish the validity of our approach. We shall
in particular consider the paradigmatic example of the long wavelength limit of the $XXX$ spin chain,
followed by the anisotropic Heisenberg model and the $gl_n$ classical magnet.
We finally consider some more complicated cases where the original quantum $R$-matrix
used to build the spin chain by coproduct
is ``twisted'' by a scalar solution of the exchange algebra. The corresponding
Hamiltonians will be discussed in general, realizing interesting formal connections between different
classical integrable models. We shall also briefly touch upon the inhomogeneous case where the
specific twist matrix will be site-dependent.
Some technical derivations will be exposed in the Appendices.

\no
Our motivation for this work is to develop a Hamiltonian approach
different in its principle from the usual Lagrangian formulation of the long wavelength limit, in
order to use in cases where the latter cannot be applied.
In our approach we start from the Hamiltonian integrability formulation (quantum R-matrix and Lax matrix)
guaranteeing a priori Liouville integrability of the classical continuous models
through a Lax matrix-classical $r$-matrix formulation, provided that some consistency checks be made.
On all known specific examples it will be checked that it yields
the same results as the Lagrangian approach.
It is indeed a key result that the Poisson structure is the same, in all cases when comparison
is available, as the canonical structure derived from the long wavelength classical Lagrangian.
This thereby validates the procedure and allows to use it in more general situations where the
Lagrangian approach may not be used, in particular as a systematic way to build more general
types of classical continuous integrable models by exploiting the richness of the algebraic
approach.

\section{The general procedure}

In this section we outline the general
procedure for obtaining a classical Lax formulation from the classical limits of the
$R$ and monodromy matrices.

\subsection{Classical limit for the $R$-matrix and the monodromy matrix}

A quantum $c$-number non-dynamical $R$-matrix obeys the quantum Yang--Baxter (YB) equation \cite{baxter}
\be
R_{12}\ R_{13}\ R_{23} = R_{23}\ R_{13}\ R_{12} \ ,
 \label{YBR}
\ee
where the labels $i = 1,2,3$ may include dependence on a complex
spectral parameter $\lambda_i$. The auxiliary spaces are in this case
loop-spaces $V_i \otimes C(\lambda_i)$, where $V_i$ are (isomorphic) finite-dimensional vector spaces.

\no
Assuming that $R$ admits an expansion (``semiclassical'') in positive power series of a parameter (usually denoted
$\hbar$) as
\be
R_{12} = 1 \otimes 1 + \hbar r_{12} + {\cal O}(\hbar^2)\ ,
\label{classlimR}
\ee
the first non-trivial term arising when we substitute this in (\ref{YBR}) is order of order two
and yields the classical YB equation
\be
[ r_{12}, r_{13}] + [r_{12}, r_{23}] + [r_{13}, r_{23}] = 0\ .
 \label{YBClR}
\ee
This is the canonically known ``classical Yang--Baxter equation''. It is not in general
the sufficient associativity condition for a classical linear Poisson bracket, except when $r$ is
non-dynamical and skew-symmetric (see e.g. \cite{STS}).
We shall hereafter limit ourselves to such situations.\footnote{
The dynamical YB equation is related to Drinfel's deformations of quantum groups, whereas
the non-skew symmetric
equation is associated to reflection algebras (see e.g. \cite{sklyanin, cherednik, Maillet}) and hence to
open spin chains which we do not consider here.}

\no
A quantum monodromy matrix $T$ is generically built as a tensor product over ``quantum spaces'' and algebraic
product over ``auxiliary space'' of representations of the YB algebra associated to $R$.
Namely, one assumes a collection operators assembled in matrices $L_{1i}$, acting on ``quantum''
Hilbert spaces labeled by $i$ and encapsulated
in a matrix ``acting'' on the auxiliary space $V_1$. For any quantum space $q$ they obey
the quadratic exchange algebra \cite{FTS, FT, tak}
\be
R_{12}\ L_{1q}\ L_{2q} =L_{2q}\ L_{1q}\ R_{12} \ ,
\label{YBRgen}
\ee
where operators acting on different quantum spaces commute.
The form of the monodromy matrix $T$ is then deduced from the co-module structure of the YB algebra
\be
T_a \equiv L_{a1}\ L_{a2}\ \ldots\ L_{aN}\
 \label{TM1}
\ee
and thus naturally obeys the same quadratic exchange algebra (\ref{YBRgen}).
In particular one can pick $L = R$,
the operators now acting on the second auxiliary space identified as ``quantum space''.
This way, one builds closed inhomogeneous
spin chains with general spins at each lattice site (labeled by $(i)$)
belonging to locally chosen representations of some Lie algebra (labeled by $i$).

\no
We now establish that $T$ has a classical limit by considering in addition the classical counterpart of
$L$, labeled by $L^c$ which then satisfies the quadratic Poisson algebra,
emerging directly as a semi-classical limit of (\ref{YBRgen}), after setting
$~{1\over \hbar} [A,\ B] \to \{ A,\ B \}$. It reads
\be
\{ L^c_a(\lambda_1),\ L^c_b(\lambda_2) \} =
[r_{ab}(\lambda_1 -\lambda_2),\ L^c_a(\lambda_1)\ L^c_b(\lambda_2)]\ .
\label{semicl0}
\ee
The quantum monodromy matrix has also a classical limit given by (see also \cite{ftbook, sklyaninlect})
\be
T^c_{a, \{i\}} =  L^c_{a 1}\ \ldots\ L^c_{a N}\ .
\label{classlimT}
\ee
The exchange algebra for $T^c$ takes the form
\be
\{T^c_a, T^c_b\} = [r_{ab},\ T^c_a\ T^c_b]\  .
\label{semicl}
\ee
This quadratic Poisson structure implies that the traces of powers of the monodromy
matrix $tr (T^c)$ generate Poisson-commuting quantities identified as classically integrable
Hamiltonians. In particular, when $T^c$ depends on a spectral parameter, the auxiliary space is a loop space
$V \otimes {\mathbb C}(\lambda)$. Performing the trace over the finite vector space yields a generating
function ${\rm tr}(T^c(\lambda))$ for classically integrable Hamiltonians obtained by series expansion
in $\lambda$.

\subsection{The long wavelength limit}

The usual presentation of the long wavelength limit,
such as can be found in \cite{fradkin, string}, is a Lagrangian one where the Poisson
structure is obtained from the standard derivation of canonical variables using a Lagrangian density.
Instead, we will present here a purely Hamiltonian version of this limit
by defining the long wavelength limit of a hierarchy of integrable quantum Hamiltonians
based on some affine Lie algebra $\hat{G}$.
We shall {\cal define a priori} the Poisson structure of the classical variables by imposing
classical integrability of the long wavelength limit of the Hamiltonian through its associated
classical Lax matrix.
We consider a $N$-site closed spin chain Hamiltonian $H$, initially
assumed to be governed by nearest-neighbour interaction that takes the form
\be
H \equiv \sum_1^N H_{l l+1}\ .
\ee
The classical, long wavelength limit, is obtained  by first defining
local quantum states as linear combinations of the base
quantum states, parametrized by a complete set of $k$ continuous variables.
The number $k$ depends on the choice of $\hat{G}$ and essentially $k = \dim(G)$.
These variables, which can be identified
as Euler angles in the simplest case of $sl(2)$,
become the classical dynamical variables once a suitable Poisson
structure is imposed. The bras and kets are denoted respectively by
$\langle  n(l, \theta_k)|$ and $|  n(l, \theta_k) \rangle$, where $l$
denotes the site index and $\theta_k$ denote the
set of $k$ angular variables. The condition of ``closed'' spin chain, essentially formulated
as $N+l \equiv l$,  imposes
periodicity or quasi-periodicity conditions on the $\theta_k$'s. We note that we assume that the
base quantum states different only by the fact that they are defined in distinct sites, hence the
frequently used notation below $|n_l\rangle$, instead of $|n(l,\th_k)\rangle$, should not be confusing.

\no
If one considers nearest-neighbor interactions (local)
then one defines the classical, but still defined in the lattice, Hamiltonian as
\be
{\cal H} \equiv \sum_1^N {\cal H}_l(t)\ ,\qq
{\cal H}_l(x, t) =  \langle  n_l| \otimes \langle n_{l+1}|\ H_{l l+1}\
| n_l\rangle\otimes |n_{l+1}\rangle\ .
\label{lwlH1}
\ee
For integrable models, we may similarly define the continuum limit of the full set of
commuting Hamiltonians. In these cases the generic Hamiltonians $H^{(n)}$
of the integrable hierarchy are obtained
{\cal directly} from the analytic
series expansion around some value $\lambda_{0}$ of the spectral parameter
of the trace of the monodromy matrix (transfer matrix) as
\be
{\rm tr} T(\lambda) \equiv \sum_{n=1}^{\infty} (\lambda -\lambda_{0})^n H^{(n)}\ .
\label{fhhjh1}
\ee
By extension, we define in this case the classical Hamiltonians as the expectation value, over the $N$ site
lattice quantum state, of $H^{(n)}$
\be
{\cal H}^{(n)}(x, t) =  \otimes_1^N \dots \langle  n_l| \otimes \langle n_{l+1}|\dots\ H^{(n)}\
\dots | n_l\rangle \otimes |n_{l+1}\rangle \dots \ .
\label{lwlH2}
\ee

\no
We next define a continuous limit and take simultaneously the thermodynamical limit in which $N \rightarrow \infty$.
Accordingly, this is achieved by identifying the lattice spacing
$\delta$ as being of order $1/N$ 
and subsequently consider only slow-varying spin configurations (the long wavelength limit proper) for
which
\be
l_i \to l(x)\ ,\qq l_{i+1}  \to l(x +\delta)\ .
\label{cont1}
\ee
In this limit, the finite ``site differences'' turn into derivatives.

\no
Given that (\ref{lwlH2}) is applied to Hamiltonians of the integrable hierarchy obtained
{\cal directly} from the series expansion of the trace of the monodromy matrix, it is
immediate that the expectation value procedure goes straightforwardly to
the full monodromy matrix $T$ (and thence to its trace over the auxiliary space which is altogether
decoupled from the quantum expectation value procedure). Accordingly, we define first a
lattice expectation value
\be
T_a = \dots \langle  n_l| \otimes \langle n_{l+1}|...\ (L_{a 1}\ L_{a 2} \ldots L_{a N})
\dots | n_l\rangle \otimes |n_{l+1}\rangle \dots \ ,
 \label{lwlL2}
\ee
which nicely factors out as
\be
T_a =  \prod_{i=1}^N \langle  n_i|  L_{ai} |n_i\rangle \ .
\label{lwlLfact2}
\ee
Assuming now that $L$ admits an expansion in powers of $\delta$ as
\be
L_{ai}= 1 + \delta l_{ai} + {\cal O}(\delta^2)\ ,
\ee
we consider the product (setting $\langle n_i|l_{ai}|n_i \rangle = l_a(x_i)$)
\be
T_a = \prod_{i=1}^N (1 + \delta l_{ai} + \sum_{n=2}^{\infty} \delta^n l^{(n)}_{ai})\ .
\ee
Expanding this expression in powers of $\d$, we get
\be
T_a = 1 + \delta \sum_i l_{ai} + \delta^2\sum_{i<j} l_{a_{i}} l_{aj} + \delta^2 \sum_{i} l^{(2)}_{ai}
+\dots \ .
\ee
These, multiple in general, infinite series of the products of local terms,
are characterized by two indices: the overall power $n$ of $\delta$, and the number $m$
of the set of indices $i$ (that is the number of distinct summation indices)
over which the series is summed. Note that, in the $T$ expansion one always has $n\geqslant m$.
The continuum limit soon to be defined more precisely, will entail the limit $\delta \to 0$ with
${\cal O}(N) = {\cal O}({1/\delta})$.
We now formulate the following {\it power-counting rule}, that is terms of the form
(for notational convenience $l_{ai}=l^{(1)}_{ai}$ below)
\be
\delta^n \sum_{i_1< i_2 <\dots i_m}l_{ai_1}^{(n_1)} ...l_{ai_m}^{(n_m)}\ ,
\qq \sum_{j=1}^m n_j =n\ ,
\ee
with $n>m$ are omitted in the continuum limit. The latter is defined by
\be
\delta \sum_{i} l_{ai} \to \int_{0}^{A} dx\ l_a(x)\
\ee
and similarly for multiple integrals. Here $A$ is the length of the continuous interval
defined as the limit of $N \delta$.
In other words, contributions to the continuum limit may only come from the terms with $n=m$
for which the power $\delta^n$ can be exactly matched by the ``scale'' factor $N^m$ of the $m$-multiple sum over
$m$ indices $i$. In particular, only terms of order one
in the $\delta$ expansion of local classical matrices $L_{ai} \equiv \langle  n_i|  L_{ai}
|n_i\rangle$ will contribute to the continuum limit.
Any other contribution acquires a scale factor $\delta^{n-m} \to 0$, when the continuum limit is taken.
This argument is of course valid term
by term in the double expansion.
Being only a weak limit argument, it always has to be checked for consistency.

\no
Let's remark that if $L$ is taken to be $R$, one naturally identifies $\delta$ with
the small parameter $\hbar$, thus identifying in some sense the classical and the continuum limits.
However, this is not required in general. It is clear to characterize separately
both notions in our discussion as
\be
{\rm classical\ limit:} && R= 1 +\hbar r\ ,
\nonumber\\
{\rm continuum\ limit:} && L = 1 + \delta l\ .
\ee
Recalling (\ref{cont1}),
the continuous limit
of $T$, hereafter denoted ${\cal T}$,
is then immediately identified
from (\ref{lwlLfact2}), as the path-ordered
exponential from $x=0$ to $x={\mathrm A}$
\be
{\cal T} = P \exp{\left (\int_{0}^{A} dx\  l(x) \right )}\ ,
\ee
where suitable (quasi) periodicity conditions on the
continuous variables $\theta_k(x)$ of the classical matrix $l(x)$,
acting on the auxiliary space $V \otimes C(\lambda)$, are assumed.
Of course the
definition of a continuous limit requires that the $L$-matrices are not
\cal{too} inhomogeneous (e.g. $L$-matrices at neighbor sites should not be too different.
This is in fact assured by the long wavelength limit assumption.

\subsection{The Lax matrix and $r$-matrix formulation}

The above identification of ${\cal T}$
also defines it as the monodromy matrix of the first order differential operator
$d/dx + l(x)$. In addition, it has been built so that to straightforwardly
generate the classical continuous limit
of the Hamiltonians in (\ref{lwlH2}) from the analytic expansion
\be
{\rm tr} ({\cal T}(\lambda)) \equiv \sum_{n=1}^{\infty} (\lambda - \lambda_{0})^n {\cal H}^{(n)} \ .
\label{fhhjh2}
\ee
We thus characterize $l(x)$ as a local Lax matrix yielding
the hierarchy of continuous Hamiltonians ${\cal H}^{(n)}$. In order for this statement to agree with the
key assumption of preservation of integrability we are now lead to
require a Poisson structure for $l$ (inducing one for the continuous dynamical variables $\theta_k(x)$)
compatible with the demand of classical integrability of the continuous Hamiltonians.
Indeed, such a structure is deduced from (\ref{semicl0}), as the ultra-local Poisson bracket
\be
\{ l_1(x, \lambda_1),\ l_2(y, \lambda_2)\}
= [r_{12}(\lambda_1 -\lambda_2),\ l_1(x, \lambda_1)+l_2(y, \lambda_2)] \delta(x-y)\ ,
\label{funda1}
\ee
where $r$ is the classical limit (\ref{classlimR}) of the $R$-matrix characterizing the exchange algebra
of the $L$-operators.
More specifically, recalling that $L_{ai} = 1 + \delta l_{ai} + {\cal O}(\delta^2)$, plugging it into (\ref{semicl0})
and assuming ultra-locality of Poisson brackets one gets
\be
\{l_{ai},\ l_{bj}\} = [r_{ab},\ l_{ai} + l_{bj}]{\delta_{ij} \over \delta}\ .
\ee
One then identifies,  in the continuum limit $\delta \to 0$, the factor
 ${\delta_{ij}/ \delta}$ with $\delta(x-y)$.
Reciprocally, it is a well known result (see, for instance L.D. Faddeev's Les Houches Lectures in
1982) that if $l(x)$ has a such an ultra-local linear Poisson bracket (\ref{funda1}) the full
monodromy matrix between $0$ and $A$
of $d/dx + l(x)$ has the quadratic Poisson bracket structure (\ref{semicl}), thereby guaranteeing
Poisson commutation of the Hamiltonians.

\no
We thus obtain a hierarchy of classically integrable, mutually Poisson commuting Hamiltonians
from the explicit computation of the monodromy matrix $t(\lambda)$ of the Lax operator $d/dx + l(x)$
as $H^{(n)} = {d^n \over d \lambda^n} t(\lambda)|_{\lambda = {\lambda_0}}$. Such Hamiltonians are
however generally highly non-local and not necessarily very relevant as physical models. We shall
thus extend our discussion to local Hamiltonians.

\subsection{The case of local spin chains}

Local spin chain Hamiltonians are more interesting, physically meaningful and
easier to manipulate. In particular, they are the most relevant objects
in connection with string theory and the AdS/CFT duality \cite{miha}. Their construction generically requires the
determination of a so-called ``regular value'' $\lambda_0$ of the spectral parameter
such that $L_{ai}(\lambda_0) \propto {\cal P}_{ai}$, where ${\cal P}$ is the
permutation operator. In this sense the expansion of
$L$ can be expressed up to an appropriate normalization factor as (see also Appendix C)
\be
L(\lambda) = f(\lambda) (1 + \delta l + {\cal O}(\delta^2)) \ .
\ee
Of course only when the auxiliary space $a$ and quantum
space $i$ are isomorphic has this ``regular value'' any relevance.
One then defines the local Hamiltonians as (denoting as usual $t(\lambda) = {\rm tr}_{a} T_a(\lambda)$)
\be
H^{(n)} = {d^n \over d \lambda^n} \ln (t(\lambda))\big |_{\lambda = {\lambda_0}} \ ,
\label{locham}
\ee
implying that they are no more such Hamiltonians expressed as linear combinations of
higher derivatives of $t(\lambda)$.
Their long wavelength limit (e.g. (\ref{lwlH1}))
is not obviously derivable from a straightforward ``diagonal''
expectation value of the $T$-matrix contrary to (\ref{lwlH2}), since
in general $\langle F(A)\rangle \neq F(\langle A \rangle)$, for any functional of
a set of operators $A$. However, we show below that this is
indeed the case due to locality properties.
Let us first focus for simplicity (but, as we shall see, without loss of generality)
on the first local Hamiltonian
\be
H^{(1)} = t(\lambda_0)^{-1}{d \over d \lambda} t(\lambda)\big |_{\lambda = {\lambda_0}}\ ,
\label{locham2}
\ee
where, $t^{-1}(\lambda_0) = {\cal P}_{12}{\cal P}_{23}\dots {\cal P}_{N-1N}$. This operator
acts exactly as a one-site shift on tensorized states, identifying of course site labels
according to the assumed periodicity, i.e. $N+1 = 1$.
(Normalization issues will be discussed in Appendix C).
Computing the expectation value of $H^{(1)}$ we obtain
\be
\langle  H^{(1)} \rangle = \langle n_1| \otimes \ldots \otimes \langle n_N| t^{-1}(\lambda_0)
{d\over d\lambda}\Big (f^N(\lambda) Tr_a\prod_{i=1}^N(1 + \delta l_{ai} + {\cal O}(\delta^2)) \Big)
|n_1\rangle \otimes \ldots \otimes |n_N\rangle \ .
\ee
One has
\be
\langle n_1| \otimes \langle n_2 | \otimes \ldots \langle n_N| t^{-1}(\lambda_0) =
\langle n_2| \otimes \langle n_3 | \otimes \ldots \langle n_1|
\ee
and of course $N+1 \equiv 1$.

\no
Taking into account the {\it power-counting rule} described in section 2.2
we obtain (see also Appendix C) that
\be
\langle H^{(1)} \rangle
= \prod_{i=1}^N \langle n_{i+1}|n_i\rangle {d\over d\lambda}
\Big (f^N(\lambda){\rm tr}_a\prod_{i=1}^N (1 + \delta \langle\ l_{ai}\ \rangle+
{\cal O}(\delta^2)) \Big )\ .
\ee
We then easily establish that in the continuum limit,
using the power counting rule and the factorized form of both the state vector
as $\langle n_1| \otimes \ldots \otimes \langle n_N|$ that the operator to
be valued over it $t^{-1}(\lambda_0) = {\cal P}_{12}{\cal P}_{23}\dots {\cal P}_{N-1N}$,
$\langle t^{-1}(\lambda_0) \rangle= \langle t(\lambda_0)\rangle^{-1}$.
We finally obtain that in the continuum limit
\be
\langle H^{(1)}\rangle = \langle t^{-1}(\lambda_0) {d\over d \lambda} t(\lambda)\big \vert_{\lambda=\lambda_0} \rangle =
\langle t(\lambda_0)\rangle^{-1}{d\over d \lambda} \langle t(\lambda)\rangle \big \vert_{\lambda= \lambda_0} =
{d \over d \lambda}(\ln \langle t(\lambda)\rangle )\big \vert_{\lambda = \lambda_0}\ .
\ee
The computation may be easily generalized along the same lines for any higher Hamiltonian.
Higher local Hamilltonians are indeed obtained from \eqn{locham},
admitting thus an expansion as
\be
H^{(n)} = t^{-1}(\lambda_0) {d^n\over d \lambda^n}t(\lambda)\big \vert_{\lambda_0}+ \mbox{polynomials}\ ,
\ee
depending only on lower order local Hamiltonians.
When computing the expectation value of such higher Hamiltonians one gets
the expectation value of $t^{-1}(\lambda_0) {d^n\over d \lambda^n}t(\lambda)\vert_{\lambda_0}$ which in
the continuum classical limit yields
\be
\langle t^{-1}(\lambda_0) {d^n\over d \lambda^n}t(\lambda)\vert_{\lambda_0}\rangle =
\langle t(\lambda_0) \rangle^{-1} {d^n\over d \lambda^n}\langle t(\lambda) \rangle \vert_{\lambda_0}\ ,
\ee
using the same arguments as in the $n=1$ case.
In addition, one obtains expectation values of the polynomials of order $k$ in the local Hamiltonians.
In this case expectation values by tensor product of local
vectors $\langle n_1| \ldots \langle n_N|$ are exactly factorized over products of $k$
local monomials $h_{i_1}\dots h_{k_k}$,
except if indices $i$ coincide (or at least overlap for multiple indices).
Locality of the lower Hamiltonians plays here a crucial role.
It is clear that such families of terms with coinciding or overlapping
indices correspond to a second ``label''
$M = k-1$ and therefore their contribution will necessarily be suppressed in the continuum limit,
with respect to the contribution of the
generic terms (non-coinciding indices) with $M =k$ by the power-counting argument.
Hence, it is consistent to conclude that in the continuum limit
\be
\langle \mbox{Polynomial\ in}\ (H^{(i)}) \rangle =  \mbox{Polynomial\ in}\ (\langle H^{(i)} \rangle)
\ee
and therefore
\be
\boxed{
\
\langle H^{(n)}\rangle  =\langle
{d^n\over d \lambda^n}\ln(t(\lambda)) \Big |_{\l=\l_0}\rangle
= {d^n\over d \lambda^n}\ln(\langle t(\lambda)\rangle) \Big |_{\l=\l_0}
\ }
\ .
\label{loccharges}
\ee
This is the final, key result in systematically establishing the classical continuum limit of integrable spin chains.
We may now apply this general procedure to all sorts of examples, starting with the
simpler applications.

\section{The XXX chain}

The XXX model Hamiltonian describing first neighbor
spin-spin interactions is given by
\be
H = {1\over 2} \sum_{j=1}^N \Big ( \sigma^x_j \sigma^x_{j+1} + \sigma^y_j
\sigma^y_{j+1} +  \sigma^z_j \sigma^z_{j+1} \Big )\ .
\ee
It is well known that when one considers the long wavelength limit
one obtains a classical $\s$-model \cite{fradkin, string}. We shall briefly review how this process works.
The coherent spin state is parametrized by the parameters $x,\ t$ via the fields $\theta, \ \varphi$ as
\be
|n(x, t)\rangle = \cos  \theta(x, t)\
e^{i \varphi(x, t) }\ | + \rangle \ +\  \sin  \theta(x, t)\ e^{-i \varphi(x, t) }\ | - \rangle\ ,
\ee
where the ranges of variables is $\th\in (0,\pi/2)$ and $\varphi\in (0,\pi)$.
One can verify the completeness relation
\be
\int d\mu( n) | n\rangle \langle n| = 1\ ,
\ee
where the integration measure is given by
\be
d \mu (n) = {4 \over \pi}\ \sin\theta \ \cos\theta\ d \theta\ d \varphi\ .
\ee
Then as was described in \cite{fradkin, string} and in subsection 2.1, one obtains a classical Hamiltonian
via the expectation value procedure by employing \eqn{lwlH1}.
The appropriate XXX 2-site Hamiltonian is
\be
H_{l l+1} \propto ({\cal P}_{l l+1} -{\mathbb I})\ ,
\label{h1}
\ee
where $\cal P$ is the permutation operator acting as
${\cal P} (a\otimes b) = b \otimes a$ for $a,\ b$ vectors in $V$.
From the definition of ${\cal H}$ we are led to compute quantities of the type
\be
\langle a| \otimes \langle b|\ {\cal P}\ |a \rangle \otimes |b\rangle = \langle a|b \rangle
\otimes \langle b |a\rangle = |\langle a|b \rangle|^2\ .
\ee
They are expressed in terms of scalar products of the form
\be
\langle  \tilde n|  n \rangle = \cos(\theta - \tilde \theta)\ \cos(\varphi - \tilde \varphi) +
i \cos(\theta + \tilde \theta)\ \sin(\varphi - \tilde \varphi)\ .
\ee
In the long wavelength limit, $| n \rangle - |\tilde n \rangle = |\delta n \rangle $,
$\tilde \theta(x)= \theta(x +\delta)$ and  $ \tilde \varphi(x) = \varphi(x + \delta)$. We conclude that
\be
H \propto \int d x \
(\theta^{'2}+\sin^2(2 \theta)\ \varphi^{'2})\ .
\label{H0}
\ee
We shall now derive the Lax representation yielding (\ref{H0}) following section 2.
The $R$-matrix for the XXX  model is \cite{yang}
\be
R(\lambda) = \lambda + i\hbar {\cal P}\ ,
\label{r1}
\ee
where for normalization issues we refer to Appendix C.
This $R$-matrix is a solution of the
quantum YB equation \cite{baxter}.
It has a consistent normalized classical limit defined as
\be
r(\lambda) = {1 \over \lambda} {\cal P}\ ,
\ee
which satisfies the classical YB equation. Alternatively,
the classical $r$-matrix may be written as
\be
r(\lambda) =  {1 \over  \lambda} \begin{pmatrix}
    {1\over 2}(\sigma^z+1) & \sigma^-  \\
    \sigma^+ & {1\over 2}(-\sigma^z+1)
\end{pmatrix} \  .
\ee
Set first
\be
L_{an}(\lambda) =R_{an}(\lambda -{i\hbar \over 2})\
\label{lan}
\ee
and demand that $L$ satisfies the fundamental algebraic relation
\be
R_{ab}(\lambda_1 -\lambda_2)\ L_{an}(\lambda_1)\ L_{bn}(\lambda_2)=
L_{bn}(\lambda_2)\ L_{an}(\lambda_1)\ R_{ab}(\lambda_1 -\lambda_2)\ ,
\label{funda0}
\ee
where as usual in the spin chain framework we call $n$ the quantum space and
$a$ the auxiliary space.
Following the general derivation of section 2 and going directly to
the continuous limit we disregard higher powers in $\delta = \hbar$ (in this case the two
small parameters are naturally identified).
We next define a ``local Lax matrix" as a mean value of $L$
on the same coherent spin state, taken solely over the quantum space
\be
\langle n| L_{an}(\lambda) | n \rangle  =   1 +i \hbar l(x,\lambda) \ ,
\ee
where
\be
l =  \begin{pmatrix}
   {1\over 2}\langle n|\sigma^z |n \rangle  & \langle n| \sigma^-  |n \rangle \\
    \langle n|\sigma^+ |n \rangle & -{1\over 2}\langle n|\sigma^z |n \rangle
\end{pmatrix}
= \ha
\begin{pmatrix}
   \cos 2 \theta (x)  & \sin 2 \theta(x)\ e^{- 2i \varphi(x)} \\
  \sin 2 \theta(x)\ e^{+ 2i \varphi(x)}& -\cos 2 \theta (x)
\end{pmatrix}\ ,
\label{llxxx}
\ee
where we have used the form of the coherent states to compute the matrix elements explicitly.
Then $l$ satisfies the
classical fundamental algebraic relation
\be
\{l_1(x, \lambda_1),\ l_2(y, \lambda_2)\} = [r_{12}(\lambda_1 -\lambda_2),\ l_1(\lambda_1)+l_2(\lambda_2)]\delta(x-y)\ .
\label{funda}
\ee
Setting $l(x, \lambda)=  \Pi/\l$
and taking into account the above algebraic relations we get
\be
\{\Pi_1,\ \Pi_2\} =  {\cal P}_{12} (\Pi_2 -\Pi_1)\delta(x-y)\ .
\ee
The parametrization in terms of the continuum parameters $\theta(x)$, $\phi(x)$
gives rise to the classical version of $sl_2$. Indeed,
parametrizing the generators of the classical current algebra as
\be
S^z =  \cos 2 \theta\ , \qq S^{\pm}= {1\over 2} \sin 2 \theta\ e^{\mp 2i \varphi}\ .
\label{par1}
\ee
we obtain from the fundamental relation that
\be
\{S^{+},S^{-}\} =  S^z\delta(x-y)\ , \qq \{S^z, S^{\pm} \} = \pm 2 S^{\pm}\delta(x-y)\ .
\label{poisson1}
\ee
The continuum parameters $\theta(x)$ and $\phi(x)$ can also be expressed in terms of canonical variables $p$ and
$q$ as
\be
\cos 2 \theta(x) = p(x)\ , \qq
\varphi(x) = q(x) ~~~\mbox{and} ~~~~\{q(x),\ p(y)\} = i \delta(x-y)\ .
\label{par2}
\ee
The $l$-matrix in \eqn{llxxx}
coincides obviously with the potential term in the Lax matrix of the classical Heisenberg model.
Precisely, one recalls that one must consider as classical Lax operator a la Zakharov--Shabat $L = d/dx + l(x)$.
The monodromy matrix for $L$ is well known now to yield the classical Hamiltonians including the first non trivial one
(see \cite{ftbook})
\be
\boxed{\
H \propto \int d x\ \left( \left({d S^z \over dx} \right)^2 +  \left({d S^x \over dx} \right)^2
+ \left({d S^y \over dx} \right)^2\right)
\ }\ .
\ee
Recalling the expressions (\ref{par1}) and substituting in the expression above we obtain the Hamiltonian (\ref{H0}),
hence the process above works consistently.

\no
Having exemplified the general construction of Section 2 to a simple system and checked the consistency
of the approach we now turn to more complicated
systems by first moving to trigonometric and elliptic $sl(2)$ $R$-matrices,
corresponding to the XXZ and XYZ spin chains.

\section{The anisotropic Heisenberg model}

Consider the generic anisotropic XYZ model with Hamiltonian
\be
H = {1\over 2} \sum_{j=1}^N \Big ( J_x \sigma^x_j \sigma^x_{j+1} + J_y\sigma^y_j
\sigma^y_{j+1} +  J_z\sigma^z_j \sigma^z_{j+1} \Big )\ .
\ee
For the following computations it is convenient to set
\be
J_{\xi} = 1 - \delta^2 a_{\xi}\ , \qq \xi \in \{x,\ y,\ z \}\ .
\ee
The Hamiltonian is written as
\be
H=  \sum_{j=1}^N{\cal P}_{j j+1} -{N\over 2} - {\delta^2 a_x \over 2} \sum_{j=1}^N \sigma^x_j \sigma^x_{j+1}
- {\delta^2 a_y \over 2} \sum_{j=1}^N \sigma^y_j \sigma^y_{j+1}
- {\delta^2 a_z \over 2} \sum_{j=1}^N \sigma^z_j \sigma^z_{j+1} \ .
\label{hdef}
\ee
The additive constant may be omitted here. Taking into account equations (\ref{h1})--(\ref{H0}), (\ref{hdef})
and keeping terms of order $\delta^2$ we get
\be
H \propto \int dx\  \Big ( \theta^{'2}+\sin^2(2 \theta)\ \varphi^{'2} +
a_x\sin^2 (2 \theta) \cos^2 (2\varphi)+
a_y\sin^2 (2 \theta) \sin^2 (2\varphi)+ a_z \cos^2( 2 \theta) \Big )\ .
\ee
This may be seen as a ``deformation'' of the classical Heisenberg Hamiltonian.
The last three terms are essentially
potential-like terms. In the special case of the XXZ model the terms with coupling constant
$a_x, a_y$ are zero, whereas in the XXX case
all potential terms vanish and one recovers the Hamiltonian (\ref{H0}).
If we now recall the parametrization (\ref{par1}),
then the expression above reduces
to the Hamiltonian of the Landau-Lifshitz model or the anisotropic classical magnet \cite{ftbook}
\be
\boxed{\
H \propto \int dx\ \left ( \left({d S_z \over dx} \right)^2 +  \left({d S_x \over dx} \right)^2 +
\left({d S_y \over dx} \right)^2 +  a_xS_x^2 +a_y
S_y^2 + a_z S_z^2\right )
\ }
\ .
\ee
We now derive the classical $l$-matrix
for the anisotropic cases. We focus in more detail on the XXZ $R$-matrix
\be
R(\lambda) = \begin{pmatrix}
    \sinh (\lambda +{i \mu \over 2}\sigma^z+{i\mu \over 2}) & \sinh (i\mu)\sigma^-  \\
    \sinh (i\mu)\sigma^+ & \sinh (\lambda -{i \mu \over 2}\sigma^z+{i\mu \over 2})
\end{pmatrix}\ .
\ee
The classical limit of the XXZ $R$-matrix, after appropriate normalization, is given as
(we divide with the constant factor $\sinh \l$)
 \be
R(\lambda) = 1 + i \mu\ r(\lambda)+{\cal O}(\mu^2)\ ,
\ee
where
\be
r(\lambda) = {1\over \sinh\lambda} \begin{pmatrix}
     ({\sigma^z \over 2} +{1 \over 2})\ \cosh \lambda & \sigma^-  \\
    \sigma^+ &  (-{\sigma^z \over 2} +{1 \over 2})\ \cosh \lambda
\end{pmatrix}\ .
\ee
The associated classical Lax operator is again obtained from $L(\lambda) = R(\lambda - {i\mu \over 2})$ as
(once again moving immediately to the continuous limit)
\be
\langle n| L(\lambda)|n \rangle  =1+ i \mu\ l(x,\lambda)+ {\cal O}(\mu^2)\ ,
\ee
where
\be
l(\lambda) &=&  {1 \over \sinh\lambda} \begin{pmatrix}
\langle n| {\sigma^z \over 2}|n\rangle \ \cosh \lambda & \langle n|\sigma^-|n\rangle  \\
\langle n|\sigma^+|n\rangle &  -\langle n|{\sigma^z \over 2}|n\rangle \ \cosh \lambda
\end{pmatrix} \non\\&=& {1\over \sinh\lambda} \begin{pmatrix}
{1\over 2}S^z \cosh \lambda & S^-  \\
S^+ &  -{1\over 2}S^z\ \cosh \lambda
\end{pmatrix}\ ,
\ee
where $S^Z,\ S^{\pm}$ are the classical generators of the current $sl(2)$ algebra realized in terms of
the angular variables in \eqn{par1}.
The continuous variables $x,y$ were omitted here for simplicity and will be from
now on whenever there is no ambiguity.

\no
Let us also briefly characterize the classical algebra underlying the model.
We set
\be
l_i(\lambda) = {\cosh \lambda \over \sinh \lambda} D_i + {1\over \sinh(\lambda)} A_i\ ,\qq
r_{12}(\lambda)
= {\cosh \lambda \over \sinh \lambda} {\cal D}_{12} + {1\over \sinh(\lambda)} {\cal A}_{12}\ .
\ee
Substituting this expressions to (\ref{funda}) and taking into account that
\be
[{\cal A}_{12},\ A_1] = - [{\cal D}_{12},\ A_2]\ ,
\ee
we end up with the following set of Poisson structures
\be
\{D_1,\ D_2\} =0\ , \quad
\{D_1,\ A_2\}= [{\cal D}_{12},\ A_2]\delta(x-y)
\ , \quad \{A_1,\ A_2\}= -[{\cal A}_{12},\ D_1]\delta(x-y)\ ,
\ee
which give rise to the $sl_2$ Poisson algebra (\ref{poisson1}).

\no
The full XYZ classical $r$-matrix also yields, through
this process, the classical Lax operator of the fully anisotropic classical Heisenberg model,
satisfying also the fundamental linear algebraic relation (\ref{funda}) (see also \cite{ftbook}).
A detailed presentation of this derivation is omitted here for the sake of brevity.

\section{The $gl_n$ classical magnet}

In this section we further extend our analysis to the case of higher rank algebras. In particular,
we study the classical limit of isotropic and anisotropic $gl_n$ type magnets.

\subsection{The isotropic case}

First consider the generic situation of the isotropic $gl_n$ quantum spin chain. The $R$-matrix
is given by the general form (\ref{r1}), where the permutation operator
is of the form
\be
{\cal P} = \sum_{i, j=1}^n e_{ij} \otimes e_{ji}\ ,\qq (e_{ij})_{kl} = \delta_{ik} \delta_{jl} \ .
\ee
The coherent state is parametrized by $n$ continuum parameters as
\be
|n(x, t)\rangle = \sum_{i=1}^n \alpha_i(x, t)\ |e_i\rangle\ ,
\ee
where $|e_i\rangle$ is the $n$ column vector with one at position $i$ and zero elsewhere. In addition
\be
\langle n|n \rangle =1 ~\Rightarrow~  \sum_{i=1}^n |\alpha_i|^2 =1\ .
\ee
Following the process described in the previous sections we end up with the
classical $r$ and $l$ operators defined as (in here $L(\l)=R(\l)$, instead of \eqn{lan})

\be
r(\lambda )= {1\over \lambda} {\cal P}\ , \qq l(\lambda) = {1\over \lambda} \sum_{i, j=1}^n
 e_{ij}\otimes  \langle n| e_{ji} |n\rangle
= {1\over \lambda} \sum_{i, j=1}^n e_{ij}\ l_{ij}\ .
\ee
The $l$-matrix satisfies the linear algebraic relation (\ref{funda}),
which clearly gives rise to the classical current-$gl_n$
exchange relations among the elements $l_{ij}(x)$. These are given by
\be
\{l_{ij}(x),\ l_{kl}(y)\} = (\delta_{il} l_{jk} -\delta_{jk} l_{il})\delta(x-y)\ .
\ee
We compute next the first local classical integral of motion starting from the spin chain Hamiltonian
\be
H^{(0)} \propto \sum_{j=1}^N {\cal P}_{j j+1}\ ,
\ee
where we have dropped from the beginning the constant compared to \eqn{h1}.
Then,
defining first the Hamiltonian density as
\be
{\cal H}^{(0)}(x) = \langle n| \otimes\langle  \tilde n|\  {\cal P}\ |n \rangle \otimes |\tilde n \rangle
= \sum_{i,j=1}^n l_{ij}(x)\ l_{ji}(x+ \delta)\ .
\ee
Expanding appropriately this, we conclude that
\be
{\cal  H}^{(0)}(x) = \sum_{i,j=1}^n l_{ij}(x)\ l_{ji}(x) - \ha
\delta^2 \sum_{i, j=1}^n {d l_{ij}(x) \over dx}{d l_{ji}(x)\over dx }\ ,
\label{iso0}
\ee
where we have dropped boundary terms by imposing appropriate boundary conditions. The first term above
is the quadratic Casimir and can be dropped.
The second term, proportional to $\delta^2$, provides, upon integration, the classical
Hamiltonian
\be
\boxed{
\
H^{(0)} \propto \int dx\  \sum_{i, j=1}^n {d l_{ij}(x) \over dx}{d l_{ji}(x)\over dx}
\ } \ .
\ee
The classical integrals of motion on the other hand are obtained from the monodromy
matrix of $l(x)$ for the generic case
along the lines described in the Appendix A (see also \cite{ftbook}). Comparing with (\ref{iso0})
they are seen to coincide.

\no
The direct computation from the classical $l(x)$ matrix is actually presented in the Appendix
for another model, but it goes along the same lines for the generalized Heisenberg model,
and is omitted here for brevity.

\subsection{The anisotropic case}

Consider now the anisotropic case.
Recall the classical $r$-matrix associated to $A_{n-1}^{(1)}$ \cite{jimbo}
\be
r(\lambda) = {\cosh (\lambda) \over \sinh (\lambda)} \sum_{i \neq j} e_{ii} \otimes e_{jj} +
{1 \over \sinh (\lambda)} \sum_{i\neq j} e^{(sgn(i-j) - (i-j){2\over n+1})\lambda} e_{ij} \otimes e_{ji}\ .
\ee
The associated classical $l$-matrix will be of the form
\be
l(\lambda) = {\cosh (\lambda) \over \sinh (\lambda)} \sum_{i \neq j} l_{jj}(x) e_{ii}+
{1 \over \sinh (\lambda)} \sum_{i\neq j} e^{(sgn(i-j) - (i-j){2\over n+1})\lambda} l_{ji}(x) e_{ij}\
\label{lani}
\ee
and satisfies the linear algebraic relation (\ref{funda}).
Take now the Hamiltonian of the deformed spin chain (see e.g. \cite{myhecke})
\be
H \propto \sum_{j} {\mathbb U}_{j j+1}\ ,
\ee
where the matrix $U$ is a representation of the Hecke algebra expressed as ($q= e^{\mu}$)
\be
{\mathbb U}= \sum_{i\neq j=1}^n \Big (e_{ij} \otimes e_{ji} - q^{-sgn(i-j)} e_{ii} \otimes e_{jj} \Big )\ .
\ee
It is convenient to rewrite it as
\be
{\mathbb U} = {\cal P} -{\mathbb I}+ \sum_{i\neq j=1}^n (1 - q^{-sgn(i-j)}) e_{ii} \otimes e_{jj}\
\ee
and also set
\be
\mu = \delta \alpha\ , \qq q^{-{\rm sgn}(i-j)} \sim 1 - {\rm sgn}(i-j) \delta \alpha + {\delta^2 \alpha^2 \over 2}\ .
\ee
The Hamiltonian in this case is basically a ``deformation'' of the isotropic case
and again the first non-trivial terms arises at order $\delta^2$.
We get that
\be
\boxed{
H \propto \int dx \left( \sum_{i, j=1}^n  {d l_{ij}(x) \over dx}{d l_{ji}(x)\over dx}
+{\alpha^2} \sum_{i\neq j=1}^n l_{ii}(x) l_{jj}(x)\ +2 a \sum_{i < j=1}^n (l_{ii}(x) l'_{jj}(x)
- l_{jj}(x) l'_{ii}(x))\right)
}\ .
\label{hani}
\ee
It clearly provides a generalization
of the Landau--Lifshitz model\footnote{Note that for $q = e^{i\mu}$ one obtains a non-Hermitian Hamiltonian.
This is not so surprising given that higher rank spin chain as well as higher $A^{(1)}_{n-1}$
affine Toda field theories with imaginary coupling are non-unitary models.
Nevertheless, the relevant physical quantities, such as spectrum excitations, exact $S$ matrices etc.
have been extensively studied.}. Note that the last term,
proportional to $a$, disappears in the case $n=2$, given that in this case $e_1 +e_2 = {\mathbb I}$.
A generalization starting from the elliptic classical $r$-matrix
can be also obtained, but is omitted here for brevity. Similarly,
this classical Hamiltonian may be again directly obtained from the
classical $l$ operator (\ref{lani}) as is described e.g. in \cite{ftbook}.

\section{Novel ``dualities'' of integrable models}

In this section we shall investigate certain
integrable ``duals'' of the XXX spin chain and its higher rank
generalizations, and we shall derive their
classical counterparts. They will be based on the
coproduct structure applied this time to $c$-number matrices.

\no
Given an initial quantum $R$-matrix, the c-number YB equation reads
\be
[R_{12}(\lambda),\ U_{1} U_2] =0\ ,
 \label{dual}
\ee
where $U$ is a particular {\cal scalar} $n \times n$ matrix.
Considering, for instance, $R$ to be the Yangian $R$-matrix,
the latter equation is valid for any $n\times n$ matrix.
In the case of the XXX matrix one may take for example the Pauli matrices to write
\be
\sigma_1^{\xi} \sigma_2^{\xi} R_{12}(\lambda) \sigma_1^{\xi} \sigma_2^{\xi} =  R_{12}(\lambda)\ .
\ee
Given the relation (\ref{dual}) we may always define a new $L$-operator $\tilde L_{12} = U_1L_{12}$, which
obviously satisfies (\ref{funda0}) as long as $L$ also satisfies it.

\no
Before we proceed with the classical limit of the models, let us
first examine the quantum local Hamiltonian arising from the $\tilde L$-matrices.
For simplicity we choose $\lambda = 0$ to be the regular value and
as usual we define this Hamiltonian as
\be
H \propto {d\over d \lambda} \log t(\lambda)\big |_{\lambda =0}\ ,
 \label{local}
\ee
where we now have
\be
t(\lambda) = {\rm tr}_0 [\tilde L_{0N}(\lambda) \ldots \tilde L_{01}(\lambda)] =
{\rm tr}_0 [U_0 R_{0N}(\lambda)
\ldots U_0 R_{01}(\lambda)]\ .
\ee
The transfer matrix at $\lambda =0$ becomes
\be
t(0) &\propto& tr_0 [U_0 {\cal P}_{0N} \ldots U_0 {\cal P}_{01}] = \ldots
= U_{N} \ldots U_{i+1}\ \Pi\ U_{i-1} \ldots \ U_1\ U_N, \\
\Pi &=& {\cal P}_{12} {\cal P}_{23} \ldots {\cal P}_{N-1 N}\ .
 \label{pp0}
\ee
Taking the derivative of the transfer matrix we find that
\be
{d t(\lambda)\over d \lambda}\Big |_{\lambda =0} \propto U_N \ldots U_{i+1}\
{d \check R_{i i+1}(\lambda) \over d \lambda}\Big |_{\lambda =0}\ \Pi \ U_{i-1} \ldots U_1\ U_N \ ,
\ee
where $\check R = {\cal P }\ R$ and we consider here the XXX $R$-matrix.
Gathering the information above we conclude that the Hamiltonian is again local and reads
\be
\boxed{H \propto \sum_{i=1}^N U_{i+1}\ {d \check R_{i i+1}(\lambda)\over d\lambda}|_{\lambda =0}\
U^{-1}_{i+1}}\ .
\label{dualh}
\ee
Note that more general
``regularity conditions'' of the form $L_{ab}(\lambda_0) = U_a{\cal P}_{ab}$, will similarly
guarantee locality of the Hamiltonians derived from (\ref{local}).

\no
An inhomogeneous generalization of this construction is obviously available by using distinct solutions
to (\ref{dual}) at each
site of the chain. One starts from a set of solutions
\be
[R_{12}(\lambda),\ U^{(i)}_{1} U^{(i)}_2] =0\ ,\qq i =1,2,\dots , N\ .
 \label{dualinh}
\ee
The quantum Hamiltonians are derived from the monodromy matrix
\be
t(\lambda) = {\rm tr}_0 [\tilde L_{0N}(\lambda) \ldots \tilde L_{01}(\lambda)] =
{\rm tr}_0 [U^{(N)}_0 R_{0N}(\lambda)
\ldots U^{(1)}_0 R_{01}(\lambda)]\ ,
\label{tr1}
\ee
by applying the co-module structure to generate an inhomogeneous transfer matrix by successive
tensoring by $U^{(i)}_0 R_{0i}(\lambda)$. Note that the same procedure (albeit with shifts
over the spectral parameter, or distinct choices of representations of the quantum
space $i$,  instead of twists by a $U^{i}$ c-number constant matrix) is used
to get inhomogeneous spin chains in many examples.

\no
Local Hamiltonians are again obtained via (\ref{local})
and following the exact steps of the proof of the homogeneous Hamiltonian above.
We end up with the generic final expression
\be
\boxed{H \propto \sum_{i=1}^N U^{(i)}_{i+1}\ {d \check R_{i i+1}(\lambda)\over d\lambda}\big |_{\lambda =0}\
(U^{(i)}_{i+1})^{-1}}\ .
\label{dualhinh}
\ee
Such Hamiltonians may be interpreted as describing spin chains in inhomogeneous backgrounds modifying
the nearest-neighbor couplings in a site-dependant pattern. Possibly interesting types of inhomogeneities
include a local defect
(one single $U_i$ distinct from $1$); a domain-like defect ($U_i = U$ for $i \leqslant i_0$ and $U_i = 1$
elsewhere) or a ``double-chain'' effect ($U_i = U$ for even $i$ and $1$ for odd $i$).
We shall not discuss the general construction of the classical limit for such chain Hamiltonians, postponing
it for future studies.

\no
We now move back to the homogeneous case.
A first remark is in order here.
It follows from (\ref{pp0}), (\ref{tr1}) that one expects to have conditions on the choice of $U^{(i)}$
and the parametrization of $|n_i\rangle$ in order to be able to define classical continuum limits.
To illustrate this point let us discuss in detail the computation of $\langle t^{-1}(0)\rangle$.
From (\ref{pp0}) one has
\be
\langle t^{-1}(0)\rangle = \otimes_i \langle n_i| ( \prod_i U_i)^{-1} \otimes_i|n_{i} \rangle =
\prod_i\langle n_{i+1}| (U_i)^{-1} |n_i\rangle\ .
\ee
It follows that if $U_i$ and $|n_i \rangle$ are such that
\be
U_i |n_i \rangle = |n_i\rangle + \delta |v_i\rangle + {\cal O}(\delta^2) \ ,
\label{condit}
\ee
where $\delta$ is the same scaling parameter as for the continuous limit of $L$
and  $|v_i\rangle$ is some vector, which can be chosen to be orthogonal to $|n_i\rangle$
without loss of generality. If this condition is fulfilled the expectation value then
has the following form
\be
\langle t^{-1}(0)\rangle = \prod_{i=1}^N (1 - \langle \delta n_i|n_i\rangle - \delta
\langle n_i| v_i \rangle + {\cal O}(\delta^2) \ ,
\ee
where the key identifications hold up to order $\delta^2$ in the discrete case. Hence
(due to the power-counting argument) exactly in the continuous limit
\be
\otimes_i \langle n_i| ( \prod_i U_i)^{-1}
\otimes_i|n_i \rangle = (\prod_i \langle n_i|  U_i | n_i \rangle)^{-1}\ .
\ee
Hence, the technical derivation of section 2.4 will hold,
yielding again $\langle t^{-1}(0)\rangle$ = $\langle t(0)\rangle^{-1} $.
It is to be expected that similar conditions will arise when considering the derivative term.

\no
We next examine the most general situation associated to the XXX model.
Consider the generic $2 \times 2$ matrix $U$ and its inverse
\be
U =\begin{pmatrix}
    a  & b \\
    c & d
\end{pmatrix}\ , \qq U^{-1} = {1\over D} \begin{pmatrix}
    d  & -b \\
    -c & a
\end{pmatrix}\ , \qq D = ad-bc\ .
\ee
Take also into account the form of $U\ \sigma^{\xi}\ U^{-1}$ we conclude that the
most general 2-site Hamiltonian after applying the $U$ transformation to XXX is given by the following form
\be
H &=& {1\over 2D} \sigma^z \otimes \Big (  (ad+bc) \sigma^z -2ab \sigma^+ +2cd \sigma^- \Big )    \non\\
  && + {1\over D} \sigma^+ \otimes  \Big ( bd \sigma^z - b^2\sigma^+ + d^2 \sigma^- \Big )
\\
  && + {1\over D} \sigma^- \otimes \Big ( -ac \sigma^z +a^2 \sigma^+ -c^2 \sigma^- \Big )\ .
\nonumber
\ee
To gain further insight we focus on particular examples.
Taking, for instance, the XXX chain and setting $U = \sigma^z$, the local Hamiltonian becomes
\be
H \propto \sum_{j=1}^N \Big (\sigma^x_j \sigma^x_{j+1} + \sigma^y_i \sigma^y_{j+1} -
\sigma^z_j \sigma^z_{j+1}\Big )\ ,
\ee
with the characteristic flip of sings in front of the $\sigma^z \otimes \sigma^z$ term.
Similarly, for $U = \sigma^{x},\sigma^y$ a minus sign in front of the
$\sigma^x \otimes \sigma^x$ and $\sigma^y \otimes \sigma^y$
terms, respectively, is attached.

\no
A classical limit can be defined for these modified Lax matrix
(recall $L(\lambda) = R(\lambda - {i\hbar \over 2})$) set also $\tilde L_{12} = U_1\ L_{12}$.
More precisely, for $U=\sigma^z$ we have, after acting from the left and right
with the coherent state:
\be
\langle n| \tilde L(\lambda) | n \rangle  =  \begin{pmatrix}
    \lambda+ {i \hbar \over 2} \cos 2 \theta  & {i \hbar \over 2} \sin 2 \theta e^{2i\varphi} \\
   - {i \hbar \over 2} \sin 2 \theta e^{-2i\varphi}                     & -\lambda+ {i \hbar\over 2} \cos 2\theta
\end{pmatrix}.
\ee
Now consider the rescaling $\th\to \hbar \th$, in the small $\hbar$ limit
and also appropriately rescale $\l\to \hbar^2 \lambda$.
This is precisely a realization of the condition [\ref{condit}) on $U$ and $| n \rangle$ discussed above.
The linear limit of the $L$ operator above becomes after setting $i \theta e^{2i\varphi} =
\psi,\ -i \theta e^{-2i\varphi} = \bar \psi $ ($\tilde L(\lambda) \sim 1 + \hbar \tilde l(\lambda)$)
\be
\langle n |\tilde  L_{a}(\lambda) |n \rangle=
 1 + i \hbar \tilde l(\lambda)\ , \qq
\tilde l(\lambda)= -2  \begin{pmatrix}
    \lambda        & \psi   \\
   \bar \psi   & -\lambda
\end{pmatrix}\ .
\ee
The $l$-matrix above is nothing else than the classical NLS Lax operator (see also \cite{izko} for lattice versions NLS).
The new spectral parameter is here defined as $\tilde \lambda = {\lambda + {i \hbar\over 2} \over \hbar^2}$
hence the critical value for $\displaystyle \tilde \lambda$ becomes
infinite in the continuum limit (in agreement with
the computations in \cite{ftbook} and Appendix A).

\no
Consider next $U = \sigma^x$ and recall the parametrization (\ref{par2}).
Taking a similar limit we get
\be
\cos 2 \theta \to \hbar p\ ,\qq \sin 2\theta \to 1\ , \qq e^{\pm 2 i \varphi} \to 1 \pm 2 i \hbar q\ ,
\qq \lambda \to \hbar^2\lambda
 \label{harmonic}
\ee
and keep only lowest order terms. The classical Lax operator takes the following form
\be
\tilde l(\lambda)= 2 \begin{pmatrix}
    -q    & -\lambda + {ip \over 2}   \\
    -\lambda  -{i p\over 2}                & q
\end{pmatrix}\ .
\ee
The latter is just the classical Lax operator for the harmonic oscillator.
Note that the classical $l$-matrices presented above are of the form: $\tilde l(\lambda)= \lambda A +B$.
Taking into account that
$\tilde l$ satisfies the linear algebraic relation (\ref{funda})
we end up with the Poisson relations
\be
\{B_1,\ B_2\} =  {\cal P} (A_1 - A_2)\delta(x-y) \ ,
\ee
which lead to the following expected canonical exchange relations
\be
\mbox{NLS model}: &&\{\psi(x), \bar \psi(y)\} =\delta(x-y)\ ,
\nonumber\\
\mbox{Harmonic oscillator}: &&\{q(x),\ p(y)\} = i\delta(x-y)\ .
\ee
We next discuss the classical limits of these ``dual'' quantum Hamiltonians. First,
consider the quantum Hamiltonian which corresponds
to the NLS model
(i.e. $U = \sigma^z$) and set
\be
{\cal H}(x) = \langle a| \otimes \langle b| \Big (\sigma^x \otimes \sigma^x + \sigma^y
\otimes \sigma^y - \sigma^z \otimes \sigma^z \Big )
|a\rangle \otimes |b \rangle\ ,
\ee
where  $\sin 2 \theta \to  2 \hbar \theta,\ \cos 2 \theta \to 1$.
The corresponding Hamiltonian becomes
\be
H = \int dx\ \theta^2(x) =\int dx\ \psi(x) \bar \psi(x)\ ,
\ee
which is just the first integral of motion of the NLS model (see \cite{ftbook} for details on the
computation of the classical integrals of motion), that is the number of particles.
Note that higher integrals of motion may be
obtained from the higher quantum Hamiltonian (higher derivatives of $\log(t(\lambda))$)
(see also Appendix B, where higher integrals of motion are also computed for NLS).
Consider now the situation where $U = \sigma^x$
\be
{\cal H}(x) \propto \langle a| \otimes \langle b|
\Big (-\sigma^x \otimes \sigma^x + \sigma^y
\otimes \sigma^y + \sigma^z \otimes \sigma^z \Big )
|a\rangle \otimes |b \rangle
\ee
and taking into account (\ref{harmonic}), the Hamiltonian density becomes
\be
{\cal H} \propto p^2(x) +{q^2(x) \over 4} \ \Rightarrow \
H \propto \int dx\ \Big (q^2(x) + {p^{2}(x)\over 4} \Big )\ ,
\ee
which is simply a classical harmonic oscillator type Hamiltonian, and coincides with the Hamiltonian
obtained directly from the classical continuum model (see Appendix A).
A similar Hamiltonian is obtained in the case where $U=\sigma^y$.
The only difference is a relative minus sign between
the $p$ and $q$ terms.
The considerations above may be generalized to the $gl_n$
case. Choose for instance
\be
&& U = \sum_{i=1}^{n-1} e_{ii} - e_{nn} \quad \mbox{also set} \non\\
&& l_{ii} \to 1\ ,  \quad  l_{nn} \to - 1\ ,  \quad  l_{in} \to -\hbar \psi_i\ ,
 \quad  l_{ni} \to \hbar \bar
\psi_i\ ,  \quad l_{ij} \to 0\ , \quad \lambda \ \to \hbar^2 \lambda \ ,
\non\\
&& i,\ j \neq n\ .
\label{dict}
\ee
Notice that above we keep only first order terms, recall also that
$l_{ij}$ are the generators of the classical $gl_n$ (see section 4).
Once again we have implemented condition (\ref{condit}).

\no
From the now standard construction
\be
\langle n |\tilde  L_{a}(\lambda) |n \rangle = 1 + i \hbar \tilde l_{a}(x, \lambda)\ ,
\ee
we conclude that the linear Lax operator in this case takes the form
\be
\tilde l(\lambda) = -i \sum_{i=1}^{n-1} \Big ( \psi_i\ e_{ni} + \bar \psi_i\ e_{in} \Big)
- \lambda \Big ( \sum_{i=1}^{n-1} e_{ii} - e_{nn} \Big )\ ,
\ee
which is just the generalized NLS Lax operator (see e.g \cite{doikoufiora} and references therein).
It is clear from (\ref{funda}) that
\be
\{\psi_i(x),\ \bar \psi_j(y)\} = \delta_{ij}\ \delta(x-y)\ .
\ee
By choosing $U$ given in (\ref{dict}) we conclude that the relevant 2-site Hamiltonian is
\be
H_2 = \sum_{k,l \neq n} e_{kl} \otimes e_{lk} - \sum_{l\neq n} e_{nl}
\otimes e_{ln} -\sum_{l\neq n} e_{ln} \otimes e_{nl} +e_{nn} \otimes e_{nn}\ ,
\ee
acting non-trivially on the sites $j,\ j+1$ of the spin chain.
Computing now the Hamiltonian density in the usual procedure, while recalling
the expansion defined in (\ref{dict}) one has
\be
{\cal H}(x) = \langle a| \otimes \langle b|\ H_{2}\ |a\rangle \otimes|b  \rangle
= \ldots \propto  \sum_{i=1}^{n-1} \psi_i(x) \bar \psi_i(x)\ .
\ee
The later provides the total number of particles of the model (see also \cite{doikoufiora})
\be
N = \int dx\ \sum_i \psi_i(x) \bar \psi(x)\ .
\ee
Similar transformations may be found in anisotropic models, however the whole analysis
is quite subtle in this situation, and will be left for future investigations.

\vskip .8 cm
\centerline{\bf Acknowledgements}

\no
J.A. wishes to thank the University of Patras for kind hospitality, and financial support during a
visit in which part of this work was completed.

\appendix

\section{Classical local Hamiltonians}

We compute the classical integrals of motion for the
classical harmonic oscillator respectively, starting from the associated
classical Lax operators. Consider the monodromy matrix
\be
T(x, y) = P \exp\left (\int_y^x dx'\ l(x', t, \lambda) \right)\ ,
\ee
satisfying the first order differential equation
\be
{d T(x,y)\over dx} = l(x)\ T(x, y)\ .
\ee
It may be expressed in the following form \cite{ftbook}
\be
T(x, y) =(1+W(x))\ e^{Z(x, y)}\ (1 +W(y))^{-1}\ ,
\label{ansatz}
\ee
where $W$ is anti-diagonal, $Z$ diagonal, and both are expanded at $\lambda \to \infty$
\be
W(x) = \sum_{m=0}^{\infty} {W^{(m)} \over \lambda^m}\ ,
Z(x) = \sum_{m=-1}^{\infty} {Z^{(m)} \over \lambda^m}\ .
\label{wz}
\ee
Our purpose is to identify the various $W^{(m)}$ and $Z^{(m)}$
and hence the associated integrals of motion.

\no
By substituting the monodromy matrix as in (\ref{ansatz}), and setting
$l(\lambda) = D + A$ ($D,\ A$
being the diagonal and anti-diagonal part of
the Lax operator) we obtain
\be
&& {d W \over dx} + 2\ D\ W + W\ A\ W  -A=0\ ,
\nonumber
\\
&& {d Z\over dx } = D + A\ W \ .
\label{equ}
\ee
Substituting expressions (\ref{wz}) in (\ref{equ}) we find
\be
&& W^{(0)} = \sigma^x\ ,
 \non\\
&& W^{(1)} = (+iq +{p\over 2}) \sigma^y\ ,
\\
&& W^{(2)} = -{1\over 2} (iq + {p\over 2})^2 \sigma^x,\ \ldots
\nonumber
\ee
and
\be
&& {d Z^{(-1)} \over dx}= -2 {\mathbb I} \ ,
\non\\
&& {d Z^{(0)} \over  dx}=0\ ,
\\
&& {d Z^{(1)} \over  dx}= -(q^2 +{p^2 \over 4}){\mathbb I}\ \ldots\ ,
\nonumber
\ee
where ${\mathbb I}$ is the $2 \times 2$ unit-matrix.
The first non-trivial integral of motion is obtained essentially from the $tr \int dx Z^{(1)}$
\be
I^{(1)} \propto \int dx (q^2(x) +{p^2(x) \over 4})\ .
\ee

\section{Higher Hamiltonians}

We focus on the computation of higher charges in the NLS
context starting from the corresponding
quantum model examined in section 6. We shall show that the quantity emerging
from the quantum higher charge is identical with the higher classical charge, that is the momentum.
This gives an illustration of the statement in Section 2, eq. \ref{loccharges}
that the second quantum local Hamiltonian
derived from the quantum $R$ matrix formulation of the spin chain, also becomes
in the continuous classical limit the second conserved quantity obtained from the monodromy matrix derived
from the classical Lax matrix $l(x)$. Hence the construction is consistent.

\no
Let us compute the quantum higher charge starting from the quantum NLS Hamiltonian
\be
H = \sum_{j} h_{j j+1}\qq , \qq h_{jj+1} =  \sigma_j^x \sigma_{j+1}^x +
\sigma_j^y \sigma_{j+1}^y -\sigma_j^z \sigma_{j+1}^x\sigma_j^z \sigma_{j+1}^y\ .
\ee
Define now the so called boost operator as (see e.g. \cite{sklyanin2} and references therein)
\be
{\cal B} = \sum_{j} j\ h_{j j+1}\ .
\ee
All higher charges in involution may be obtained via the boost operator ${\cal B}$ as
follows
\be
H^{(n+1)} = [{\cal B},\ H^{(n)}]\ .
\label{boost}
\ee
So the next charge one obtains via (\ref{boost}) is of the form
\be
H^{(2)} \propto \sum_{j} [h_{j j+1},\ h_{j+1 j+2}]\ .
\ee
The three site quantum higher Hamiltonian is then
\be
h^{(2)}&=& [h \otimes {\mathbb I},\ {\mathbb I} \otimes h]\ , \qq
h = \sigma^x \otimes \sigma^x + \sigma^y \otimes \sigma^y -
\sigma^z \otimes \sigma^z\ .
\ee
It is now straightforward to show that
\be
 h^{(2)} = \sigma^x \otimes (\sigma^z \otimes \sigma^y+ \sigma^y \otimes \sigma^z) -
\sigma^y \otimes (\sigma^x \otimes \sigma^z + \sigma^z \otimes \sigma^x) -
\sigma^z \otimes (\sigma^y \otimes \sigma^x - \sigma^x \otimes \sigma^y)\ .
\ee
Define now the Hamiltonian density as
\be
{\cal H}^{(2)}(x) = \langle n| \otimes \langle n| \otimes \langle n|\
h^{(2)}\ | n \rangle \otimes | n \rangle \otimes | n \rangle\ .
\ee
Recalling the identifications: $\sin 2\theta \to 2\hbar \theta,~~\cos 2\theta \to 1$, we have
\be
\langle n| \sigma^x |n \rangle \to 2 \hbar\theta\ \sin 2 \varphi\ , \qq
\langle n| \sigma^y |n \rangle \to 2\hbar \theta\ \cos 2 \varphi\ ,
\qq \langle n| \sigma^z |n \rangle \to 1
\ee
and taking into account the first non trivial contribution (${\cal O}(\delta)$),
after expanding the difference
operators between neighbor sites
$\theta_{i+1} - \theta_i \to \theta(x+\delta) - \theta(x)$ (same for $\varphi_i$),
we conclude that (
recall also $i \theta e^{2i\varphi} = \psi,\ -i \theta e^{-2i\varphi} = \bar \psi $)
\be
{\cal H}^{(2)} =4 \theta^2\ \varphi' \ \propto\ \psi(x) \bar \psi'(x)- \psi'(x) \bar \psi(x)\ .
\ee
And indeed the second  conserved quantity is the momentum of NLS i.e.
\be
H^{(2)} \propto \int dx\ \Big ( \psi(x) \bar \psi'(x)- \psi'(x) \bar \psi(x) \Big )\ .
\ee
It is clear that similar computations can be done for higher charges and for other models,
but these are beyond the intended scope
of the present work. We simply focus here on a simple example the NLS case to
further illustrate the consistency of our approach.

\section{Local spin chains: normalization factor}

A delicate normalization issue arises in the considerations of section 2.4.
Superficially the assumption of ``regular limit'' of the $L$
matrix $L_{ai}(\lambda_0) \equiv {\cal P}_{ai}$ clashes with the assumption
of ``semiclassical limit'' $L_{ai} = 1 \otimes 1 + \delta l_{ai}$.
In fact one is considering two different
normalizations of the same initial $R$ matrix, yielding respectively
the semiclassical $L_{cl}$ matrix and the regular $L_r$ matrix. They will differ
in the simplest case by an overall c-number factor as $L_{\rm cl} = \delta (\lambda - \lambda_0)^{-1} L_r$.

\no
The transfer matrix $t$ yielding ${d \log t(\lambda) \over d \lambda}\big |_{\lambda = \lambda_0}$,
with $t^{-1}(\lambda_0) = {\cal P}_{12}{\cal P}_{23}\dots {\cal P}_{N-1 N}$
is obtained from application of the co-module structure to $L_r$.
Whenever an overall normalization factor $f(\lambda)$
is applied to $L$ the ``new'' $T$ matrix acquires an overall factor $f(\lambda)^N$ and
the Hamiltonian $H^{(1)}$
is shifted by a trivial identity operator $ N {d \over d \lambda} \log f(\lambda)|_{\lambda =\lambda_0}$.
One can then regularize the whole construction by suitable substraction of this
infinite identity
operator whilst keeping quantum integrability.

\no
The remaining problem in this case is the issue of having to consider the
continuous limit of the product
$\prod_1^N ( 1 \otimes 1 + \delta l_{1i})$ around the regular
value $\lambda_0$ since it would appear from the definition of
 $L_{\rm cl} = (\lambda - \lambda_0)^{-1} L_r$
that the generic term of the infinite product is then singular. But in fact one is always dealing
with formal series expansion \rm{around} this value, or equivalently integrals on an arbitrarily
small contour around but not touching this value.
It is understood that evaluations of derivatives of the $\ln$ of the transfer matrix
at point $\lambda = \lambda_0$
generically mean extraction of the leading term in formal series expansion.
The $\delta \rightarrow 0$ limit
is thus always defined, and the continuous limit thus computed will
generate, as shown in the various examples studied, the correct Hamiltonians, providing a
definite check of consistency of the procedure.

\end{document}